# Cloud Versus Local Processing in Distributed Networks

Abdulaziz M. Alqarni, *Member, IEEE,* and Thomas G. Robertzzi, *Fellow, IEEE*

*Abstract*—**A method for evaluating the relative performance of local, cloud and combined processing of divisible (i.e. partitionable) data loads is presented. It is shown how to do this in the context of Amdahl's law. A single level tree (star) network operating under each of three fundamental scheduling policies is used as an example. Applications include mobile computing, cloud computing and signature searching.**

## I. Introduction

Data processing is an important part of the fast-growing computer and communication technologies. The expanding multiple processor technologies requires effective and efficient data processing scheduling. It is important to have a complete understanding to avoid economic waste, inaccurate future system designs and a lack of technological improvements. There are two types of processing. The first one is serial processing, which cannot be divided and processed simultaneously. The second type and the focus of this paper is parallel processing. We consider a type of parallel processing where loads of data are divisible among processors, which allow efficient and effective parallel processing. Such divisible load scheduling also known as divisible load theory utilizes linear mathematical models [1-6]. DLT has many advantages such as easy computation, a schematic language and equivalent network element modeling. DLT can be implemented in many applications such as intelligent sensors, image signal processing and large data bases. Given expanding sensor information collection abilities, there is a requirement for execution time prediction tools. DLT provides scalable and tractable models to be used as an accurate prediction tool [7].

This paper compares local and cloud computing as well as combining both methods. Local computing simply means all the computation occurs on a local network. However, cloud computing means all the data computation and storage happens in the cloud via the internet [8]. A combination of local and cloud computing will be tested and compared to the results of using only one computing method. Three single level tree models will be used to test those processing methods. Also, the testing included both heterogeneous and homogeneous networks for those models. Finally, the finish time and the optimal load distribution will be calculated for each case.

Section II discusses the related theory for this paper. The notation used in this paper can be found in section III. The formulation of this problem and solution appears in section IV.

The results are analyzed in section V. The conclusion is in section VI.

## II. Related Theory

### A. Divisible Load Scheduling

Divisible load scheduling consists of two steps: load distribution and load processing. The data is usually distributed from one or more processors to multiple processors and processed in parallel. An optimal scheduling provides the minimum finish time. In DLT, there are no precedence relations between the data, which allow data to be divided among a number of processors and links. Also, network architectural issues related to parallel and distributed computing can be solved implementing DLT. Dividing the load equally among the processors does not take different computer and communication link speeds, the scheduling policy and the interconnection network into account and that leads to suboptimal solutions. However, an optimal solution can be found using divisible load scheduling theory, which provides the required mathematical tools. Furthermore, the solution can be improved by integrating Amdahl's and other speedup laws.

### B. Amdahl's Law

Amdahl's law is an accurate formula to calculate the speedup of the execution of a task with fixed data size [9,10]. For multiple processors networks, Amdahl's law can be used as a prediction tool for the theoretical speedup in parallel computing. For example, suppose a program finish time using a single processor is 10 hours. If the part of the program that cannot be parallelized takes one hour to execute and the part that can be parallelized takes the remaining 9 hours ($p = 0.9$) of execution, then the minimum execution time cannot be less than that critical one hour. Hence, the theoretical speedup is limited to at most 10 times.

A task can be split up into two parts: a part that cannot be parallelized *(1-f)* and a parallelized part *f*. The execution time of the whole task is $T_s$. The parallelized part execution time can be improved by the factor *s* which is the speedup of the parallelized part [7]. Consequently, the theoretical execution time *T(s)* of the whole task after the improvement can be calculated using the following formula:

$$T(s) = (1\text{-}f)\,T_s + \frac{f}{s}T_s \qquad (1)$$

Amdahl's law gives the theoretical speedup of the execution of the whole task *at fixed workload W,* which yields

$$S_{Amdahl}(s) = \frac{T_s W}{T(s)W} = \frac{T_s}{T(s)} = \frac{1}{(1-f)+\frac{f}{s}} \qquad (2)$$

The problem formulation, solution and results are discussed in the following sections. The divisible load speedup expressions are included for three fundamental load distribution protocols in a single level tree network [1]. Heterogeneous and homogenous networks are included for each protocol. Moreover, the integrated speedup formulas will be used for the cases of local, cloud, and a combination of local and cloud computing.

## III. Notation

The following notation is used in this paper:

| | |
|---|---|
| $n$ | The number of processors. |
| $\omega_0$ | The inverse of the computing speed of the source node. |
| $\omega_c$ | The inverse of the computing speed of the cloud. |
| $\omega_i$ | The inverse of the computing speed of the $i^{th}$ processor. |
| $z_i$ | The inverse of the link speed of $i^{th}$ link. |
| $T_{cm}$ | Communication intensity constant: the entire load is transmitted in $z_i T_{cm}$ seconds over the $i^{th}$ link. |
| $T_{cp}$ | Computing intensity constant: the entire load is processed in $\omega_i T_{cp}$ seconds by the $i^{th}$ processor. |
| $S_{DLT}(n)$ | The speedup with $n$ processors in the systems using a DLT model. |
| $S_{DLT_{homo}}(n)$ | The speedup with $n$ homogeneous processors and links in the systems using a DLT model. |
| $f$ | The fraction of load that is parallelizable. |
| $(1-f)$ | The fraction of load that is serial. |
| $S_p$ | The parallel system speedup. |
| $S_{Amdahl}$ | The overall system speedup. |
| $\alpha_i$ | The load fraction assigned to the $i^{th}$ link-processor pair. |
| $T_{f,m}$ | The finish time for m processors. |

## IV. Problem Formulation and Solution

Optimal load distribution is a key to calculate an efficient finish time $T_{f,m}$. Here $T_{f,m}$ indicates the time it takes for a single level tree consisting of the source node and m children nodes to finish processing the entire load. The finish time (make span) in this section will be used as a measurement tool to compare local and cloud processing. Also, a combination of both processing methods will be tested and compared. Local computing means all the computation occurs on a local network (single level tree network). Cloud computing means all the data computation and storage happens in the cloud via the internet (the cloud is a single node connected to the root node via a link). A combination of both processing methods means the data processing is done partially on a local network and partially in the cloud. In other words, the processing load will be divided and distributed between the cloud and the local processors. When combining both processing methods for a single level tree network, the cloud processor will be modeled as one of the children nodes. The calculation will be done on three single level tree models; sequential load distribution model, simultaneous distribution, staggered start model, and simultaneous distribution, simultaneous start model.

Divisible load modeling and speedup expressions have been developed for a variety of multi-processor interconnection topologies such as buses, stars, multi-level tree networks, meshes, hypercubes and other networks. Also, they have been developed for different load distribution policies such as sequential load distribution and concurrent load distribution with simultaneous or staggered start. Amdahl's Law can be modified and used to calculate the entire network speedup including serial and parallel data. The speedup of a divisible load model is implemented as the parallel part of the system replacing the "s" in Amdahl's law. This integration allows other factors to be included in Amdahl-like Laws such as interconnection topology, load distribution policy and the relative difference in computation and communication intensity and speed [7].

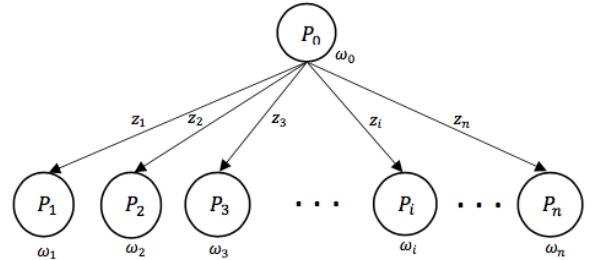

Figure 4.1. Single level tree network

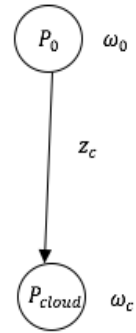

Figure 4.2. Cloud computing

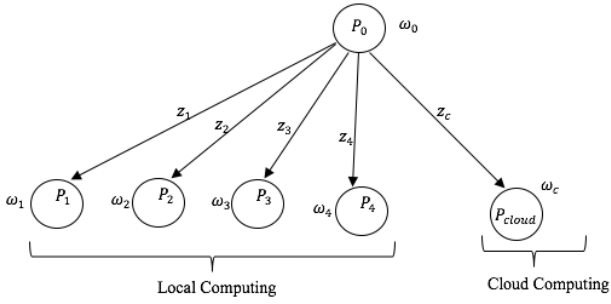

Figure 4.3. Combination of local and cloud computing

Figure 4.1 shows a single level tree network [1]. The load is distributed from the root node to the children nodes. As shown in the figure above, $\omega_i$ is the $i$th processor's inverse computing speed and $z_i$ is the $i$th link's inverse link speed. It is assumed that computation takes more time than communication which means that the product of the inverse link speed and the communication intensity constant is smaller than the product of the inverse computing speed and computation intensity constant. The case of pure cloud computing is shown in Fig. 4.2 and the case of local and cloud computing is illustrated in Fig. 4.3.

All the values used in this section are included in Table 4.1. The problem formulation, solution and results are discussed in this section of this paper. Also, the divisible load speedup expressions are included for three different single level tree network protocols. Heterogeneous and homogenous networks are included for each protocol. Moreover, the integrated speedup formulas will be used for the cases of local and cloud computing. Finally, the finish time (make span) and the optimal load distribution will be calculated for each case.

| Network Type | $n$ | $\omega_0$ | $\omega_c$ | $\omega_1, \omega_2, \omega_3, \omega_4$ | $z_c, z_1, z_2, z_3, z_4$ | $T_{cp}$ | $T_{cm}$ |
|---|---|---|---|---|---|---|---|
| Heterogeneous system | 4 | 2 | 0.3 | 4,5,6,7 | 0.1,1.5,2,2.5,3 | 2 | 1 |
| Homogeneous system | 4 | 3 | 0.3 | 3,3,3,3 | 0.1,0.1,0.1,0.1,0.1 | 2 | 1 |

Table 4.1. Values of parameters used in the calculation

### A. Sequential Load Distribution model:

The timing diagram for a single level tree network with a sequential load distribution is shown in Figure 4.4 [1]. It illustrates the communication and computation parts of this protocol. Sequential load distribution means the source node distributes load to one child node at a time. The child node starts processing as soon as it begins receiving the assigned load. In other words, the child node does not have to wait until the assigned load is completely received to start computing. To achieve the optimal speedup, all the nodes have to finish processing at the same time [2,3].

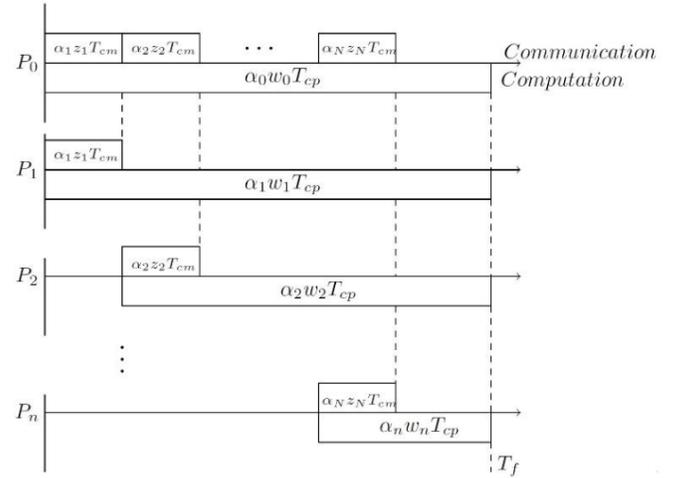

Figure 4.4. Timing Diagram for a single level tree network with a sequential load distribution

One has, since all processors stop computing at the same time:

$$\alpha_0 \omega_0 T_{cp} = \alpha_1 \omega_1 T_{cp} \qquad (3)$$

$$\alpha_0 = \frac{\omega_1}{\omega_0} \alpha_1 = \frac{1}{k_1} \alpha_1$$

where, $k_1 = \frac{\omega_0}{\omega_1}$

$$\alpha_{i-1} \omega_{i-1} T_{cp} = \alpha_{i-1} z_{i-1} T_{cm} + \alpha_i \omega_i T_{cp} \qquad (4)$$

$$\alpha_i = \frac{\omega_{i-1} T_{cp} - z_{i-1} T_{cm}}{\omega_i T_{cp}} \alpha_{i-1} = q_i \alpha_{i-1} = (\prod_{l=2}^{i} q_l) \times \alpha_1$$

where, $q_i = \frac{\omega_{i-1} T_{cp} - z_{i-1} T_{cm}}{\omega_i T_{cp}}$

$$\alpha_0 + \alpha_1 + \alpha_2 + \cdots + \alpha_m = 1 \qquad (5)$$

Using Equations (3, 4) and (5) with unit load provides the following equation:

$$[\frac{1}{k_1} + 1 + \sum_2^m (\prod_{l=2}^i q_l)]\alpha_1 = 1$$

$$\alpha_1 = \frac{1}{[\frac{1}{k_1} + 1 + \sum_2^m (\prod_{l=2}^i q_l)]} \qquad (6)$$

One can use Equation (6) and the values in table 4.1 to find the finish time using the following equation:

$$T_{f,m} = \alpha_0 \omega_0 T_{cp} = \frac{1}{k_1} \alpha_1 \omega_0 T_{cp} = \frac{1}{1 + k_1 [1 + \sum_2^m (\prod_{l=2}^i q_l)]} \omega_0 T_{cp} \qquad (7)$$

All the results for the Sequential Load Distribution model are shown in table 4.2 for the parameters of Table 4.1

| Sequential Load Distribution | Local Processing | Cloud Processing | Combination of Local and Cloud Processing |
|---|---|---|---|
| Heterogeneous Network | 1.84 | 0.522 | 0.463 |
| Homogeneous Network | 1.225 | 0.545 | 0.446 |

Table 4.2. The finish time results for the Sequential Load Distribution model

The speedup of a divisible load model of a single level tree network with a sequential load distribution parallel facility with $n$ processors is $S_{DLT}(n)$. $S_{DLT}(n)$ can be calculated using the following equation [1]:

$$S_{DLT}(n) = 1 + k_1 \left[1 + \sum_{i=2}^{n}\left(\prod_{l=2}^{n} q_l\right)\right] \qquad (8)$$

Where  $q_i = (\omega_{i-1}\ T_{cp} - z_{i-1}\ T_{cm})/\omega_i\ T_{cp}$

and        $k_1 = \omega_0/\omega_1$

In the homogeneous case, the inverse processor speed for each processor is the same. Moreover, the inverse link speeds are all the same. In this case, $S_{DLT}(n)$ in Equation (8) can be modified and simplified to calculate the homogeneous network speedup $S_{DLT_{homo}}(n)$ using the following equation:

$$S_{DLT_{homo}}(n) = 1 + \frac{\omega_0}{\omega}\left[\frac{1-(1-\sigma)^n}{\sigma}\right] \qquad (9)$$

where: $\sigma = zT_{cm}/\omega T_{cp}$

Amdahl's Law can be modified to calculate the speedup of the entire network including both serial and parallel facilities using the following equation [7]:

$$S_{Amdahl} = \frac{1}{(1-f) + \frac{f}{S_{DLT}(n)}} \qquad (10)$$

To test and compare the speedup levels for different networks, the values of the parameters listed in Table 4.1 are implemented in equations (8) and (9). Once the speedup is calculated using equations (8) and (9) the results can be implemented in equation (10) to calculate the overall speedup of the system. All the results for the heterogeneous processors are shown in Table 4.3 and Figure 4.5. Also, homogeneous processors are tested, and the results are shown in Table 4.4 and Figure 4.6.

| f | 1-f | f/sp local | f/sp cloud | f/sp comb. | Ss local | Ss cloud | Ss comb. |
|---|---|---|---|---|---|---|---|
| 0.000 | 1.000 | 0.000 | 0.000 | 0.000 | 1.000 | 1.000 | 1.000 |
| 0.100 | 0.900 | 0.046 | 0.013 | 0.012 | 1.057 | 1.095 | 1.097 |
| 0.200 | 0.800 | 0.092 | 0.026 | 0.023 | 1.121 | 1.211 | 1.215 |
| 0.300 | 0.700 | 0.138 | 0.039 | 0.035 | 1.193 | 1.353 | 1.361 |
| 0.400 | 0.600 | 0.184 | 0.052 | 0.046 | 1.275 | 1.533 | 1.547 |
| 0.500 | 0.500 | 0.230 | 0.065 | 0.058 | 1.369 | 1.769 | 1.793 |
| 0.600 | 0.400 | 0.276 | 0.078 | 0.069 | 1.479 | 2.091 | 2.130 |
| 0.700 | 0.300 | 0.322 | 0.091 | 0.081 | 1.607 | 2.556 | 2.625 |
| 0.800 | 0.200 | 0.368 | 0.104 | 0.093 | 1.760 | 3.286 | 3.418 |
| 0.900 | 0.100 | 0.414 | 0.117 | 0.104 | 1.944 | 4.600 | 4.899 |
| 1.000 | 0.000 | 0.460 | 0.130 | 0.116 | 2.172 | 7.667 | 8.643 |

Table 4.3. Heterogeneous processors testing results for sequential load distribution

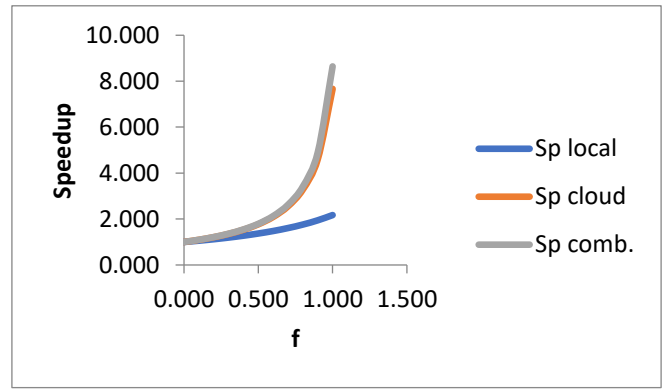

Figure 4.5. Heterogeneous processors results for sequential load distribution

| f | 1-f | f/sp local | f/sp cloud | f/sp comb. | Ss local | Ss cloud | Ss comb. |
|---|---|---|---|---|---|---|---|
| 0.000 | 1.000 | 0.000 | 0.000 | 0.000 | 1.000 | 1.000 | 1.000 |
| 0.100 | 0.900 | 0.020 | 0.009 | 0.007 | 1.086 | 1.100 | 1.103 |
| 0.200 | 0.800 | 0.041 | 0.018 | 0.014 | 1.189 | 1.222 | 1.228 |
| 0.300 | 0.700 | 0.061 | 0.027 | 0.021 | 1.314 | 1.375 | 1.387 |
| 0.400 | 0.600 | 0.082 | 0.036 | 0.028 | 1.467 | 1.571 | 1.592 |
| 0.500 | 0.500 | 0.102 | 0.045 | 0.035 | 1.661 | 1.833 | 1.869 |
| 0.600 | 0.400 | 0.122 | 0.055 | 0.042 | 1.914 | 2.200 | 2.262 |
| 0.700 | 0.300 | 0.143 | 0.064 | 0.049 | 2.258 | 2.750 | 2.864 |
| 0.800 | 0.200 | 0.163 | 0.073 | 0.056 | 2.753 | 3.667 | 3.904 |
| 0.900 | 0.100 | 0.184 | 0.082 | 0.063 | 3.525 | 5.500 | 6.129 |
| 1.000 | 0.000 | 0.204 | 0.091 | 0.070 | 4.900 | 11.000 | 14.248 |

Table 4.4. Homogeneous processors results for sequential load distribution

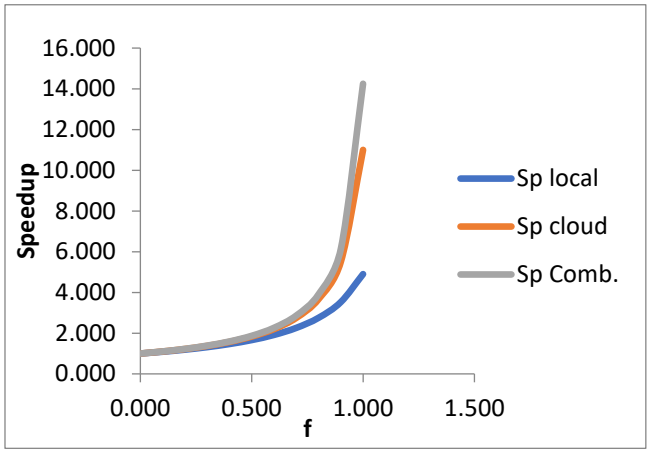

Figure 4.6. Homogeneous processors results for sequential load distribution

## B.  *Simultaneous Distribution, Staggered Start model*

The timing diagram for a single level tree network with a simultaneous distribution, staggered start is shown in Figure 4.7 [1]. The communication and computation parts of this protocol are included in the figure. The second model involves simultaneous distribution of load which means the source node distributes the divisible load to all children nodes simultaneously. In this model, a child node can start load processing only after receiving the entire assigned load from the source node. To achieve the optimal speedup, all the nodes have to finish processing at the same time.

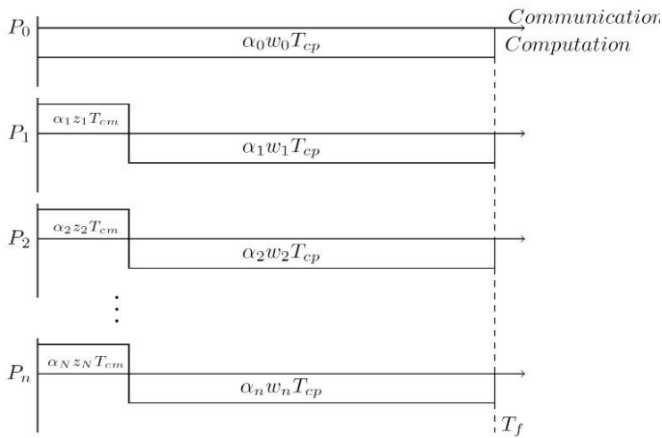

Figure 4.7. Timing Diagram for a single level tree network with a simultaneous load distribution and staggered start

One has, since all processors stop computing at the same time:

$$\alpha_0 \omega_0 T_{cp} = \alpha_1 \omega_1 T_{cp} + \alpha_1 z_1 T_{cm}  \qquad (11)$$

$$\alpha_0 = \frac{\omega_1 T_{cp} + z_1 T_{cm}}{\omega_0 T_{cp}} \ \alpha_1 = \frac{1}{k_1} \alpha_1 \qquad \text{where, } k_1 = \frac{\omega_0 T_{cp}}{\omega_1 T_{cp} + z_1 T_{cm}}$$

$$\alpha_{i-1} \omega_{i-1} T_{cp} + \alpha_{i-1} z_{i-1} T_{cm} = \alpha_i \omega_i T_{cp} + \alpha_i z_i T_{cm} \qquad (12)$$

$$\alpha_i = \frac{\omega_{i-1} T_{cp} + z_{i-1} T_{cm}}{\omega_i T_{cp} + z_i T_{cm}} \ \alpha_{i-1} = q_i \alpha_{i-1} = (\prod_{l=2}^{i} q_l) \times \ \alpha_1 \ \text{ where,}$$

$$q_i = \frac{\omega_{i-1} T_{cp} + z_{i-1} T_{cm}}{\omega_i T_{cp} + z_i T_{cm}}$$

$$\alpha_0 + \alpha_1 + \alpha_2 + \cdots + \alpha_m = 1 \qquad (13)$$

Using Equations (11, 12) and (13) provides the following equation:

$$[\frac{1}{k_1} + 1 + \sum_2^m (\prod_{l=2}^i q_l)] \alpha_1 = 1$$

$$\alpha_1 = \frac{1}{[\frac{1}{k_1} + 1 + \sum_2^m (\prod_{l=2}^i q_l)]} \qquad (14)$$

Using Equation (14) and the values in table 4.1 to find the finish time using the following equation:

$$T_{f,m} = \alpha_0 \omega_0 T_{cp} = \frac{1}{k_1} \alpha_1 \omega_0 T_{cp} = \frac{1}{1 + k_1 [1 + \sum_2^m (\prod_{l=2}^i q_l)]} \ \omega_0 T_{cp} \qquad (15)$$

All the results for the Simultaneous Distribution, staggered start model are shown in table 4.5 below.

| Simultaneous Distribution, Staggered Start | Local Processing | Cloud Processing | Combination of Local and Cloud Processing |
|---|---|---|---|
| Heterogeneous Network | 1.767 | 0.596 | 0.501 |
| Homogeneous Network | 1.213 | 0.628 | 0.445 |

Table 4.5. The finish time results for the Simultaneous Distribution, Staggered Start model

The speedup of a divisible load model of a single level tree network with a simultaneous load distribution and staggered start parallel facility with $n$ processors is $S_{DLT}(n)$. $S_{DLT}(n)$ can be calculated using the following equation [1]:

$$S_{DLT}(n) = 1 + \omega_0 T_{cp} \sum_{i=1}^{n} \frac{1}{(\omega_i T_{cp} + z_i T_{cm})} \qquad (16)$$

In the homogeneous case, the inverse link speeds are all the same. Also, the inverse processor speed for each processor is the same. In this case, $S_{DLT}(n)$ in Equation (16) can be modified and simplified to calculate the homogeneous network speedup $S_{DLT_{homo}}(n)$ using the following equation:

$$S_{DLT_{homo}}(n) = 1 + k \times n \qquad (17)$$

$$\text{where: } \quad k = \omega_0 T_{cp} / (\omega T_{cp} + z T_{cm})$$

Amdahl's Law can be modified to calculate the speedup of the entire network including both serial and parallel facilities using Equation (10). To test and compare the speedup levels for different networks, the values of the parameters listed in Table 4.1 are implemented in equations (16) and (17). Once the speedup is calculated using equations (16) and (17) the results can be implemented in equation (10) to calculate the overall speedup of the system. The testing results of heterogeneous networks are shown in Table 4.6 and Figure 4.8. Also, Homogeneous processors are tested, and the results are shown in Table 4.7 and Figure 4.9.

| f | 1-f | f/sp local | f/sp cloud | f/sp comb. | Ss local | Ss cloud | Ss comb. |
|---|---|---|---|---|---|---|---|
| 0.000 | 1.000 | 0.000 | 0.000 | 0.000 | 1.000 | 1.000 | 1.000 |
| 0.100 | 0.900 | 0.044 | 0.015 | 0.013 | 1.059 | 1.093 | 1.096 |
| 0.200 | 0.800 | 0.088 | 0.030 | 0.025 | 1.126 | 1.205 | 1.212 |
| 0.300 | 0.700 | 0.132 | 0.045 | 0.038 | 1.201 | 1.343 | 1.356 |
| 0.400 | 0.600 | 0.177 | 0.060 | 0.050 | 1.288 | 1.516 | 1.538 |
| 0.500 | 0.500 | 0.221 | 0.074 | 0.063 | 1.387 | 1.741 | 1.777 |
| 0.600 | 0.400 | 0.265 | 0.089 | 0.075 | 1.504 | 2.043 | 2.104 |
| 0.700 | 0.300 | 0.309 | 0.104 | 0.088 | 1.642 | 2.474 | 2.579 |
| 0.800 | 0.200 | 0.353 | 0.119 | 0.100 | 1.808 | 3.133 | 3.330 |
| 0.900 | 0.100 | 0.397 | 0.134 | 0.113 | 2.011 | 4.273 | 4.699 |
| 1.000 | 0.000 | 0.442 | 0.149 | 0.125 | 2.265 | 6.714 | 7.979 |

Table 4.6. Heterogeneous processors results for simultaneous load distribution and staggered start

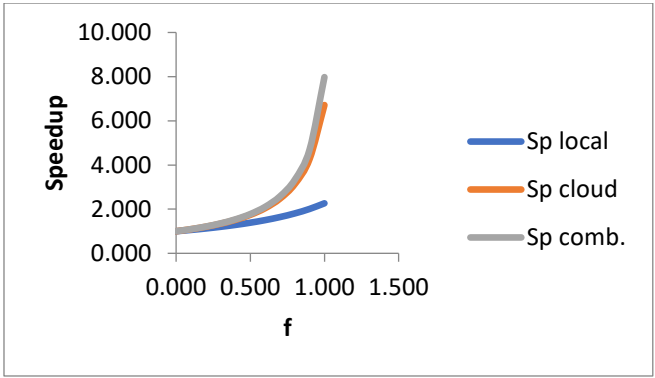

Figure 4.8. Heterogeneous processors results for simultaneous load distribution and staggered start

| f | 1-f | f/sp local | f/sp cloud | f/sp comb. | Ss local | Ss cloud | Ss comb. |
|---|---|---|---|---|---|---|---|
| 0.000 | 1.000 | 0.000 | 0.000 | 0.000 | 1.000 | 1.000 | 1.000 |
| 0.100 | 0.900 | 0.020 | 0.010 | 0.007 | 1.087 | 1.098 | 1.102 |
| 0.200 | 0.800 | 0.041 | 0.021 | 0.015 | 1.190 | 1.218 | 1.227 |
| 0.300 | 0.700 | 0.061 | 0.031 | 0.022 | 1.314 | 1.367 | 1.385 |
| 0.400 | 0.600 | 0.081 | 0.042 | 0.030 | 1.468 | 1.558 | 1.588 |
| 0.500 | 0.500 | 0.101 | 0.052 | 0.037 | 1.663 | 1.811 | 1.862 |
| 0.600 | 0.400 | 0.122 | 0.063 | 0.044 | 1.917 | 2.161 | 2.250 |
| 0.700 | 0.300 | 0.142 | 0.073 | 0.052 | 2.263 | 2.680 | 2.842 |
| 0.800 | 0.200 | 0.162 | 0.084 | 0.059 | 2.762 | 3.526 | 3.858 |
| 0.900 | 0.100 | 0.182 | 0.094 | 0.067 | 3.542 | 5.154 | 6.001 |
| 1.000 | 0.000 | 0.203 | 0.104 | 0.074 | 4.936 | 9.571 | 13.507 |

Table 4.7. Homogeneous processors results for simultaneous load distribution and staggered start

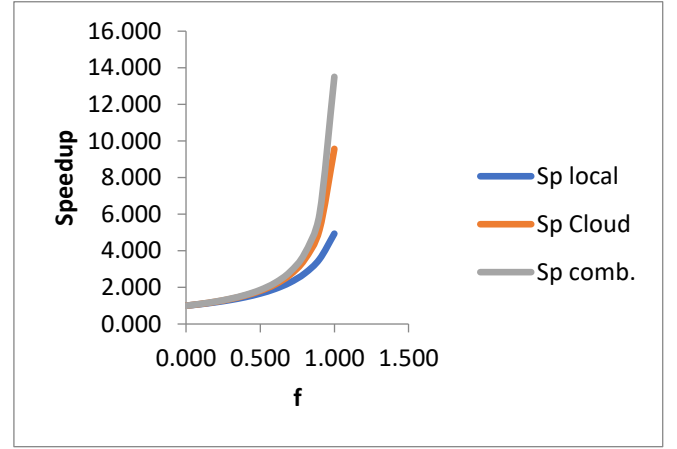

Figure 4.9. Homogeneous processors results for simultaneous load distribution and staggered start

## C. Simultaneous Distribution, Simultaneous Start model

The timing diagram for a single level tree network with a simultaneous distribution, simultaneous start is shown in Figure 4.10 [1]. The figure includes communication and computation parts of this protocol. This model involves simultaneous distribution of load which means the source node distributes the divisible load to all children nodes simultaneously. In this model, a child node can start load processing as soon as it begins to receive its assigned load from the source node. In other words, the child node does not have to wait until the assigned load is completely received to start computing. All the nodes have to finish processing at the same time to achieve the optimal speedup.

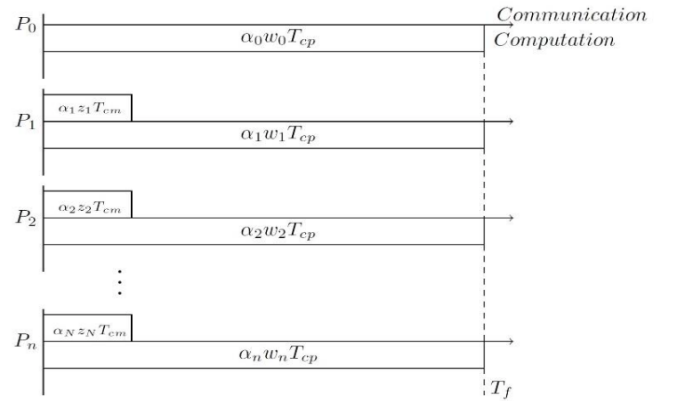

Figure 4.10. Timing diagram of a single level tree network with a simultaneous distribution, simultaneous start

One has, since all processors stop computing at the same time:

$$\alpha_0 \omega_0 T_{cp} = \alpha_1 \omega_1 T_{cp} \qquad (18)$$

$$\alpha_0 = \frac{\omega_1}{\omega_0}\alpha_1 = \frac{1}{k_1}\alpha_1 \qquad \text{where, } k_1 = \frac{\omega_0}{\omega_1}$$

$$\alpha_{i-1}\omega_{i-1}T_{cp} = \alpha_i \omega_i T_{cp} \qquad (19)$$

$$\alpha_i = \frac{\omega_{i-1}T_{cp}}{\omega_i T_{cp}} \; \alpha_{i-1} = q_i \alpha_{i-1} = (\prod_{l=2}^{i} q_l) \times \; \alpha_1$$

where, $q_i = \frac{\omega_{i-1}}{\omega_i}$

$$\alpha_0 + \alpha_1 + \alpha_2 + \cdots + \alpha_m = 1 \tag{20}$$

Using Equations (18, 19) and (20) provides the following equation:

$$[\frac{1}{k_1} + 1 + \sum_{2}^{m}(\prod_{l=2}^{i} q_l)]\alpha_1 = 1$$

$$\alpha_1 = \frac{1}{[\frac{1}{k_1} + 1 + \sum_{2}^{m}(\prod_{l=2}^{i} q_l)]} \tag{21}$$

Using Equation (21) and the values in table 4.1 to find the finish time using the following equation:

$$T_{f,m} = \alpha_0 \omega_0 T_{cp} = \frac{1}{k_1} \alpha_1 \omega_0 T_{cp} = \frac{1}{1 + k_1[1 + \sum_{2}^{m}(\prod_{l=2}^{i} q_l)]} \; \omega_0 T_{cp} \tag{22}$$

All the results for the Simultaneous Distribution, simultaneous start model are shown in table 4.10.

| Simultaneous Distribution, Simultaneous Start | Local Processing | Cloud Processing | Combination of Local and Cloud Processing |
|---|---|---|---|
| Heterogeneous Network | 1.584 | 0.522 | 0.436 |
| Homogeneous Network | 1.2 | 0.545 | 0.428 |

Table 4.8. The finish time results for the Simultaneous Distribution, Simultaneous Start model

The speedup of a divisible load model of a single level tree network with a simultaneous distribution, simultaneous start parallel facility with $n$ processors is $S_{DLT}$ $(n)$. $S_{DLT}$ $(n)$ can be calculated using the following equation [1]:

$$S_{DLT} \; (n) = 1 + \omega_0 \; \sum_{i=1}^{n} (\frac{1}{\omega_i}) \tag{23}$$

For the system with homogeneous processors, the inverse processing speed for each processor is the same and the inverse link speeds are the same. In this case, $S_{DLT}$ $(n)$ in Equation (23) can be modified and simplified to calculate the homogeneous network speedup $S_{DLT_{homo}}(n)$ using the following equation:

$$S_{DLT_{homo}}(n) = 1 + k \times n \quad \text{where} \quad k = \omega_0/\omega \tag{24}$$

Amdahl's Law can be modified to calculate the speedup of the entire network including both serial and parallel facilities using Equation (10). To test and compare the speedup levels for different networks, the values of the parameters listed in Table 4.1 are implemented in equations (23) and (24). Once the speedup is calculated using equations (23) and (24) the results can be implemented in equation (10) to calculate the overall speedup of the system. The testing results of heterogeneous networks are shown in Table 4.9 and Figure 4.11. Also, Homogeneous processors are tested, and the results are shown in Table 4.10 and Figure 4.12.

| f | 1-f | f/sp local | f/sp cloud | f/sp comb. | Ss local | Ss cloud | Ss comb. |
|---|---|---|---|---|---|---|---|
| 0.000 | 1.000 | 0.000 | 0.000 | 0.000 | 1.000 | 1.000 | 1.000 |
| 0.100 | 0.900 | 0.040 | 0.013 | 0.011 | 1.064 | 1.095 | 1.097 |
| 0.200 | 0.800 | 0.079 | 0.026 | 0.023 | 1.137 | 1.211 | 1.216 |
| 0.300 | 0.700 | 0.119 | 0.039 | 0.034 | 1.221 | 1.353 | 1.362 |
| 0.400 | 0.600 | 0.159 | 0.052 | 0.045 | 1.318 | 1.533 | 1.550 |
| 0.500 | 0.500 | 0.198 | 0.065 | 0.057 | 1.432 | 1.769 | 1.796 |
| 0.600 | 0.400 | 0.238 | 0.078 | 0.068 | 1.567 | 2.091 | 2.137 |
| 0.700 | 0.300 | 0.278 | 0.091 | 0.079 | 1.730 | 2.556 | 2.636 |
| 0.800 | 0.200 | 0.318 | 0.104 | 0.091 | 1.932 | 3.286 | 3.441 |
| 0.900 | 0.100 | 0.357 | 0.117 | 0.102 | 2.187 | 4.600 | 4.951 |
| 1.000 | 0.000 | 0.397 | 0.130 | 0.113 | 2.519 | 7.667 | 8.826 |

Table 4.9. Heterogeneous processors results for simultaneous load distribution and simultaneous start

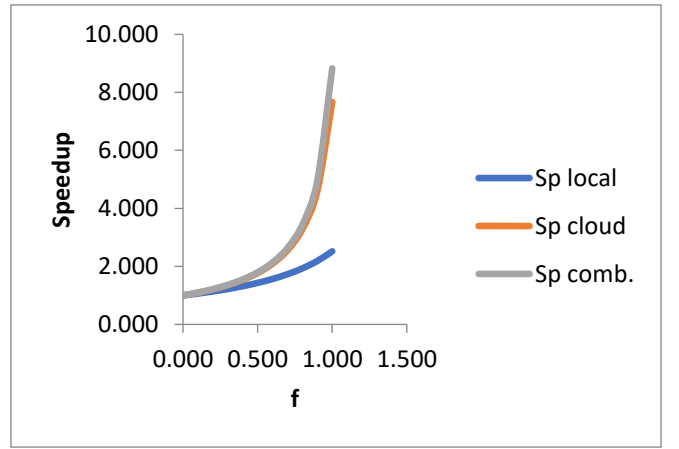

Figure 4.11. Heterogeneous processors results for simultaneous load distribution and simultaneous start

| f | 1-f | f/sp local | f/sp cloud | f/sp comb. | Ss local | Ss cloud | Ss comb. |
|---|---|---|---|---|---|---|---|
| 0.000 | 1.000 | 0.000 | 0.000 | 0.000 | 1.000 | 1.000 | 1.000 |
| 0.100 | 0.900 | 0.020 | 0.009 | 0.007 | 1.087 | 1.100 | 1.103 |
| 0.200 | 0.800 | 0.040 | 0.018 | 0.013 | 1.190 | 1.222 | 1.230 |
| 0.300 | 0.700 | 0.060 | 0.027 | 0.020 | 1.316 | 1.375 | 1.389 |
| 0.400 | 0.600 | 0.080 | 0.036 | 0.027 | 1.471 | 1.571 | 1.596 |
| 0.500 | 0.500 | 0.100 | 0.045 | 0.033 | 1.667 | 1.833 | 1.875 |
| 0.600 | 0.400 | 0.120 | 0.055 | 0.040 | 1.923 | 2.200 | 2.273 |
| 0.700 | 0.300 | 0.140 | 0.064 | 0.047 | 2.273 | 2.750 | 2.885 |
| 0.800 | 0.200 | 0.160 | 0.073 | 0.053 | 2.778 | 3.667 | 3.947 |
| 0.900 | 0.100 | 0.180 | 0.082 | 0.060 | 3.571 | 5.500 | 6.250 |
| 1.000 | 0.000 | 0.200 | 0.091 | 0.067 | 5.000 | 11.000 | 15.000 |

Table 4.10. Homogeneous processors results for simultaneous load distribution and simultaneous start

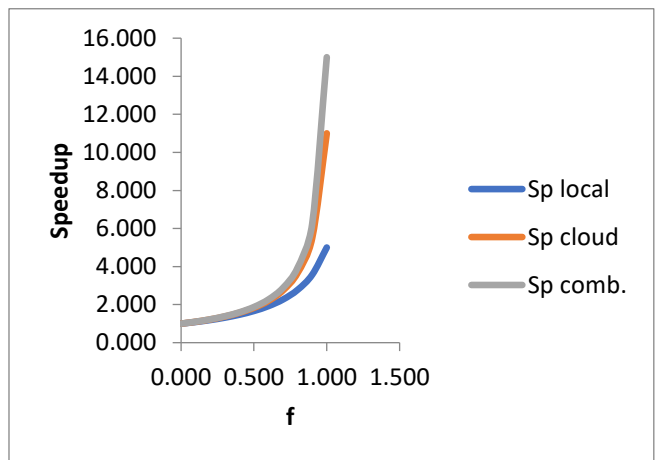

Figure 4.12. Homogeneous processors results for simultaneous load distribution and simultaneous start

## V. Analysis

After finishing all the calculation, the results were analytically reviewed. Throughout working on this paper, it became clear that the results depend on the implemented parameters. In other words, a certain set of parameters will provide a specific answer that a different set of parameters wouldn't provide. Based on the previous statement all the following analysis are based on the work and results mentioned in the previous section of this paper. It is meant to be illustrative and show that type of calculations and tradeoffs are possible.

The results of the first model sequential load distribution in table 4.3 and table 4.4 shows clearly that the speedup values for the system with homogeneous processors are higher than the values of the system with heterogeneous processors for the set of parameters in table 4.1. The reason is the processing speed for the homogeneous processors equals the highest processing speed among the heterogenous processors in the specified parameters. In other words, the homogenous system has a higher computation power than the heterogeneous system. The results of the second model Simultaneous Distribution, Staggered Start and the third model Simultaneous Distribution, Simultaneous Start also demonstrate this statement. Moreover, the results of the optimal load distribution and the finish time support this statement. The finish time results for all three models are in Table 4.2, 4.5 and 4.8. For all three models, systems with homogeneous processors have a smaller finish time results than a system with heterogeneous processors. However, that is not true for the cloud computing model. It is solved as a heterogeneous model since the cloud processor is faster than the source node.

Analyzing the results of local processing speedup for different network topologies shows clearly that simultaneous distribution models have higher speedup values than the sequential distribution model. This is because the source node in the sequential distribution model send the assigned divisible loads to the children nodes one at a time. In other words, the children nodes ranked lower in the sequence must wait for a considerable length of time. However, the children nodes can all start receiving load near the starting time in the simultaneous distribution model. The local processing results in Table 4.3, 4.4, 4.6, 4.7, 4.9 and 4.10 support this analysis. However, in the cases of cloud processing and a combination of local and cloud processing, the sequential load distribution model provided better speedup than simultaneous distribution staggered start since there is a delay in processing the load in the staggered start model for these parameters.

After comparing sequential load distribution with the simultaneous distribution topologies, the next step is to compare the simultaneous distribution topologies. The simultaneous distribution topologies are simultaneous distribution, staggered start model and simultaneous distribution, simultaneous start model. The simultaneous start model results in Table 4.9 and 4.10 shows higher speedup values than the staggered start model. The reason is all the nodes can start processing as soon as they begin receiving their assigned load. Unlike the staggered start model, children nodes do not have to wait for the assigned load to be completely received to start processing.

Finally, all the results in the first part of the solution section shows the influence of the size of the parallelizable load ($f$) on the speedup values of different network topologies used in this chapter. Overall, the model has a higher speedup values when the value of ($f$) increases. That can be seen clearly in Figures 4.5, 4.6, 4.8, 4.9, 4.11 and 4.12.

## VI. Conclusion and Future Work

Amdahl's law is an effective tool to evaluate the performance of parallel systems by calculating the speedup. The speedup value can be utilized to compare the performances of different system topologies. Amdahl's law and divisible load modeling were integrated to evaluate different single level tree topologies. divisible load modeling was used because it provides tractable and realistic models. Also, it allows precise mathematical analysis.

To summarize:

- It is possible to model and compare local, cloud and combined computing in a divisible load framework.
- It is possible to incorporate Amdahl's Law into this framework.
- Trade-offs and optimizations between the three approaches can be considered.

Such mathematical analysis leads the way in designing capable computer systems and predicting their performance.

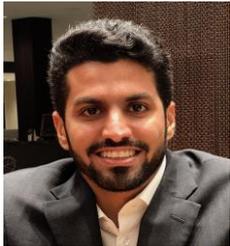

**Abdulaziz M. Alqarni** received a B.S.E.E. from Western Michigan University, Kalamazoo, MI, in 2014, M.S. degree in electrical engineering from Loyola Marymount University, Los Angeles, CA, and the Ph.D. from the Department of Electrical Engineering and Computer Engineering in Stony Brook University, Stony Brook, NY, in 2020.

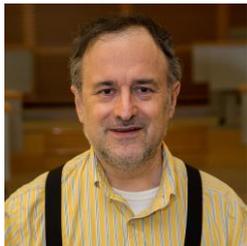

**Thomas G. Robertazzi** received the Ph.D. from Princeton University, Princeton, NJ, in 1981 and the B.E.E. from the Cooper Union, New York, NY in 1977.

He is presently a Professor in the Dept. of Electrical and Computer Engineering and an Affiliate Professor in the Dept. of Applied Mathematics and Statistics at Stony Brook University, Stony Brook N.Y. He has published in the areas of parallel processing, telecommunications, switching, queuing and Petri networks.

He has authored, co-authored or edited six books in the areas of performance evaluation, scheduling and network planning. He is a Fellow of the IEEE.